\documentclass[twocolumn,secnumarabic,amssymb, nobibnotes, aps, prd, superscriptaddress]{revtex4-2}

\usepackage{amsmath,mathrsfs,amssymb,amsfonts,amsthm,mathtools}
\usepackage{graphicx,color}
\usepackage{subcaption}
\usepackage{booktabs}
\usepackage{multirow}
\usepackage{physics}
\usepackage[T1]{fontenc}
\usepackage{bm}
\usepackage{dsfont}
\usepackage{empheq}
\usepackage{hyperref}
\hypersetup{
    colorlinks=true,
    linkcolor=blue,
    citecolor=blue,
    urlcolor=blue
}

\newcommand{\nn}{\nonumber}

\begin{document}
\title{Worldline-Induced Transparency}

\author{Arash Azizi}
	\affiliation{Texas A\&M University, College Station, TX 77843}

\begin{abstract}
We show that the Unruh response can be interferometrically suppressed or restored in a single Unruh--DeWitt detector whose center-of-mass is prepared in a coherent superposition of two uniformly accelerated worldlines. The two paths remain physically disjoint; the detector is read out in a \emph{path-erasing} basis so that no which-path information is revealed. If the detector's energy gap is path dependent during the interaction, the branch amplitudes for first-order excitation become operationally indistinguishable and therefore add coherently. With appropriate tuning---matching the gap-to-acceleration ratios of the two branches and choosing a single relative phase---the conditional first-order excitation amplitude cancels, while reversing the phase restores the response. We derive these conditions in two complementary formalisms and interpret the mechanism as a relativistic analogue of electromagnetically induced transparency, which we term \emph{worldline-induced transparency} . We also treat finite switching times explicitly and quantify how imperfect matching produces a residual signal, yielding a tolerance window rather than an idealized infinitely sharp condition. 
\end{abstract}

\maketitle


\section{Introduction}
\begin{figure}[t]
\centering
\includegraphics[width=\columnwidth]{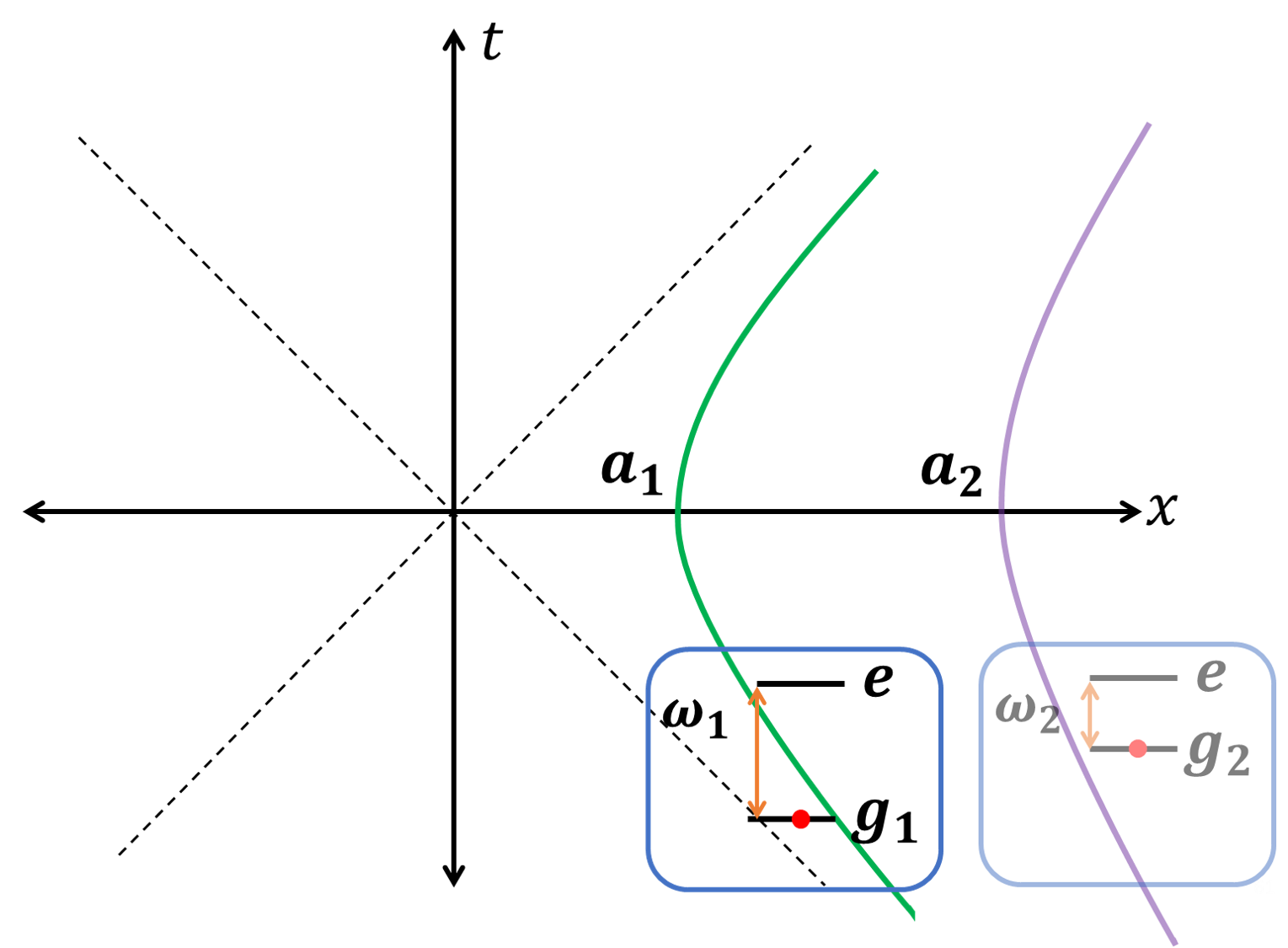}
\caption{Conceptual schematic of WIT. A single detector is coherently split between two uniformly accelerated, disjoint worldlines ($a_1,a_2$). The excited state is \emph{common} ($\ket{e}$), while the ground states are \emph{branch dependent} ($\ket{g_1},\ket{g_2}$) via distinct gaps ($\omega_1,\omega_2$). The lighter purple arm indicates a smaller branch weight in the superposition ($|\alpha_1|$ vs.\ $|\alpha_2|$).}
\label{fig:concept}
\end{figure}

The Unruh effect---the thermal response of an accelerated detector in the Minkowski vacuum---is a cornerstone of quantum field theory in curved spacetime~\cite{Unruh1976,Hawking1975,Fulling1973,Birrell_Davies1982,Wald1994,Crispino2008}. For a detector constrained to a \emph{single} worldline, the response is kinematic and universal. Here we ask a complementary, intrinsically quantum question: how does this response change when the detector's center of mass is prepared in a coherent superposition of two accelerated worldlines? We consider a single detector delocalized between two physically disjoint, uniformly accelerated paths. The branches are not recombined in spacetime; instead, the detector is read out in a \emph{path-erasing} basis, i.e., a measurement basis that does not reveal which worldline the detector followed. Under such a readout, the conditional excitation amplitude is a coherent sum of two branch amplitudes and can therefore exhibit destructive interference. We refer to the resulting interference-based suppression as \emph{worldline-induced transparency} (WIT) and comment on its relation to electromagnetically induced transparency (EIT) as an interpretive analogy~\cite{Fleischhauer2005,Harris1997,Boller1991}. Figure~\ref{fig:concept} sketches the setting: a \emph{single} detector is coherently split between two uniformly accelerated, spatially disjoint worldlines (with proper accelerations $a_1$ and $a_2$), while the final measurement is performed in a superposition basis so that the two alternatives remain operationally indistinguishable; in this conditional (path-erasing) output, the two branch excitation pathways interfere just like the two arms of an interferometer (with unequal branch weights indicated schematically by different arm intensities).

The central theoretical ingredient is a conservative modification of the standard Unruh--DeWitt (UDW) detector model~\cite{Unruh1976,Einstein100,Colosi2009Rovelli}: during the interaction, the detector's internal energy gap is conditioned on the branch, so that the ground-state splitting can differ between the two worldlines. This is the only additional structure we assume beyond a single pointlike detector coupled locally to the field; in particular, we do not introduce auxiliary detectors, nonlocal couplings, or new field degrees of freedom. Branch-conditioned internal dynamics is routine in quantum-control platforms. In trapped ions, state-dependent forces generate controlled internal--motional correlations that directly realize branch-dependent phases and splittings~\cite{Leibfried2003RMP}. In superconducting circuits, the transition frequency of flux-tunable qubits can be rapidly and coherently tuned by external biasing, providing a natural route to branch-conditioned gaps~\cite{Blais2021RMP_cQED}. Phase-stable superpositions of spatially separated matter-wave paths with coherent recombination are standard in atom interferometry~\cite{Cronin2009RMP}, including Raman-pulse beam splitters that coherently split and recombine atomic wave packets~\cite{Kasevich_Chu1991PRL}. Stern--Gerlach interferometry provides a complementary route in which the splitting and recombination are explicitly spin dependent~\cite{Machluf2013NatComm}. On the theory side, detector models with quantized or delocalized center-of-mass degrees of freedom have been developed in relativistic quantum-information settings~\cite{Stritzelberger2021,Gale2023RelUDW}, and interferometric strategies have been proposed for analogue-Unruh detection in Bose--Einstein condensates~\cite{Gooding2020Interferometric}. More broadly, superconducting platforms have been developed as a route to probe quantum-vacuum phenomena (including Unruh/Hawking-related analogues)~\cite{Nation2012Nori}, while analogue black-hole experiments in condensates have reported quantum Hawking radiation and correlations in the emitted quasiparticles~\cite{Steinhauer2016}.

WIT is, at heart, an \emph{interference effect}: the Unruh response is not merely a rate associated with a trajectory, but an amplitude that can add or cancel when alternative worldlines are made indistinguishable by measurement. In our setting, the path-erasing readout projects onto a superposition of the branch ancilla states, so the conditional excitation channel functions like selecting an output port of an interferometer. The key consequence is that, at leading order in the detector--field coupling, the excitation probability in that conditional port is controlled by \emph{coherent} addition of two branch amplitudes rather than by an incoherent sum. As a result, the Unruh response can be \emph{suppressed} (a ``dark'' output) or \emph{restored/enhanced} (a ``bright'' output) by tuning a single relative phase between the branches.

We derive simple analytic rules for when this cancellation occurs. First, the two branches must populate the \emph{same} one-particle sector of the field, which enforces a matching of the dimensionless gap-to-acceleration ratios,
\begin{align}
\frac{\omega_1}{a_1}=\frac{\omega_2}{a_2},
\end{align}
so that the postselected branch contributions can interfere rather than living in orthogonal mode subspaces. Second, once this matching holds, a single relative phase of the branch weights selects the interference sign, yielding either destructive interference (WIT) or constructive interference (restoration). We establish these conditions in two complementary ways---directly in the Unruh-mode decomposition and independently in a Minkowski plane-wave representation---making explicit how mode orthogonality and operator rephasings constrain when interference is physically meaningful.

Moreover, finite interaction times do not change the mechanism; they simply replace exact mode matching by an overlap requirement set by the Fourier width of the switching window. We treat this explicitly in Sec.~\ref{sec:finite}, and for Gaussian switching we obtain a closed-form first-order amplitude and an explicit tolerance window (Sec.~\ref{subsec:finite_gauss}). The rest of the paper is organized as follows: in Sec.~\ref{sec:model} we introduce the branch-conditioned detector model and derive the adiabatic cancellation/restoration rules in the Unruh-mode basis; in Sec.~\ref{sec:plane_wave} we provide an independent verification in the Minkowski plane-wave representation; in Sec.~\ref{sec:finite} we analyze finite-duration switching and the resulting natural bandwidth, and in Sec.~\ref{sec:conclusion} we conclude by summarizing the interference picture and outlining extensions. The derivation of the Gaussian-switched amplitude is given in Appendix~\ref{app:gauss_derivation}.

\section{Model and realization} \label{sec:model}

We work in $1+1$ dimensions with a massless scalar field and set $c=1$. We consider a single Unruh--DeWitt (UDW) detector whose center-of-mass is coherently delocalized between two uniformly accelerated worldlines in the right Rindler wedge, with proper accelerations $a_1>0$ and $a_2>0$. A path ancilla $\hat{\wp}$ with orthonormal eigenstates $\{\ket{a_1},\ket{a_2}\}$ labels the two branches. During the interaction window the detector’s internal gap is conditioned on $\hat{\wp}$,
\begin{align}
\omega(\hat{\wp})\,\ket{a_k}=\omega_k\,\ket{a_k},
\qquad (k=1,2).
\end{align}
We take the excited state to be common to both branches, denoted $\ket{e}$, while the ground state depends on the branch, $\{\ket{g_1},\ket{g_2}\}$. The initial state is
\begin{align}
\ket{\Psi_i}
=\big(\alpha_1\ket{g}_{a_1}+\alpha_2\ket{g}_{a_2}\big)\ket{0_M},
\label{eq:initial_state}
\end{align}
where $\ket{g}_{a_k}\equiv\ket{a_k}\otimes\ket{g_k}$ and $\ket{0_M}$ is the Minkowski vacuum.

The interaction Hamiltonian in the interaction picture is written in terms of the detector proper time $\tau$. We employ derivative coupling with a smooth switching function $s(\tau)$,
\begin{align}
H_I(\tau)
= g\,s(\tau)\,\frac{d}{d\tau}
\Big[\Phi_{\mathrm{RTW}}(u(\tau))+\Phi_{\mathrm{LTW}}(v(\tau))\Big]\;m(\tau),
\label{eq:interaction_H}
\end{align}
where $g$ is the detector--field coupling constant and $s(\tau)$ is a dimensionless window that localizes the interaction in proper time (Sec.~\ref{sec:finite} treats finite-time profiles explicitly). For a massless scalar field in $1+1$ dimensions one may decompose the field into right- and left-traveling sectors,
\begin{align}
\Phi(t,x)=&\Phi_{\mathrm{RTW}}(u)+\Phi_{\mathrm{LTW}}(v),
\nn\\
u=&t-x, \qquad v=t+x,
\end{align}
so that the interaction samples the two independent chiral components along the worldline through
$\Phi_{\mathrm{RTW}}(u(\tau))+\Phi_{\mathrm{LTW}}(v(\tau))$.

The ancilla-dependent monopole operator is
\begin{align}
m(\tau)
= \sigma_-(\hat{\wp})\,e^{-i\,\omega(\hat{\wp})\,\tau}
+\sigma_+(\hat{\wp})\,e^{+i\,\omega(\hat{\wp})\,\tau},
\label{eq:monopole}
\end{align}
i.e.\ the usual two-level UDW dipole operator in the interaction picture, with branch-conditioned splitting $\omega(\hat{\wp})$. The lowering and raising operators are
\begin{align}
\sigma_-(\hat{\wp})
&=\sum_{k=1}^2 \ket{g_k}\bra{e}\otimes\ket{a_k}\bra{a_k},\nn\\
\sigma_+(\hat{\wp})
&=\sigma_-^\dagger(\hat{\wp}),
\end{align}
so that $\sigma_-(\hat{\wp})\ket{e}\otimes\ket{a_k}=\ket{g_k}\otimes\ket{a_k}$ and $\sigma_-(\hat{\wp})\ket{g_j}=0$.

Along branch $k$, we parametrize the uniformly accelerated trajectory using light-cone coordinates evaluated on that branch,
$u_k(\tau)\equiv u\big(t_k(\tau),x_k(\tau)\big)$ and
$v_k(\tau)\equiv v\big(t_k(\tau),x_k(\tau)\big)$. For uniform acceleration in the right Rindler wedge one may choose
\begin{align}
-a_k\,u_k(\tau)=e^{-a_k\tau},
\qquad
a_k\,v_k(\tau)=e^{+a_k\tau},
\label{eq:trajectory_param}
\end{align}
so that the proper-time derivative acts as
\begin{align}
\frac{d}{d\tau}
=(-a_k u_k)\,\partial_{u_k}
=(a_k v_k)\,\partial_{v_k}.
\label{eq:dtau_lightcone}
\end{align}
The derivative coupling therefore introduces the standard dimensionless ratios $\omega_k/a_k$ in the excitation amplitudes.

We expand the chiral field components in Unruh modes~\cite{Unruh1976,UnruhWald1984}. Keeping only the creation-operator parts,
\begin{align}
\Phi^-_{\mathrm{RTW}}(u)
&= \int_{-\infty}^{+\infty} d\Omega \;
f(-\Omega)\,(-u)^{-i\Omega}\,A^\dagger_{\Omega}, \nn\\
\Phi^-_{\mathrm{LTW}}(v)
&= \int_{-\infty}^{+\infty} d\Omega \;
f(\Omega)\,v^{-i\Omega}\,B^\dagger_{\Omega},
\label{eq:unruh_modes}
\end{align}
with
\begin{align}
f(\Omega)=\frac{e^{-\pi\Omega/2}}{\sqrt{8\pi\,\Omega\,\sinh(\pi\Omega)}}.
\end{align}
The operators $A^\dagger_{\Omega}$ and $B^\dagger_{\Omega}$ create single Unruh quanta labeled by the dimensionless Rindler/Unruh frequency $\Omega$ in the right- and left-moving sectors, respectively; states with different labels are orthogonal (e.g.\ $A^\dagger_{\Omega}\ket{0_M}\perp A^\dagger_{\Omega'}\ket{0_M}$ for $\Omega\neq \Omega'$). The superscript ``$-$'' indicates that only creation operators appear.

To first order in perturbation theory,
\begin{align}
\ket{\Psi_f}
=-\frac{i}{\hbar}\int_{-\infty}^{\infty} d\tau\, H_I(\tau)\ket{\Psi_i},
\end{align}
which yields
\begin{align}
\ket{\Psi_f}
=&-\frac{ig}{\hbar}\sum_{k=1}^2 \alpha_k
\int_{-\infty}^{\infty} d\tau\,s(\tau)\,m(\tau)\nn\\
&\times \frac{d}{d\tau}\Big[\Phi_{\mathrm{RTW}}(u_k)+\Phi_{\mathrm{LTW}}(v_k)\Big]\,
\ket{g}_{a_k}\ket{0_M}.
\label{eq:dyson}
\end{align}
In the long-interaction (adiabatic) limit, boundary terms induced by $\partial_\tau$ vanish and the remaining integrals are understood in the distributional sense; the finite-time case is treated explicitly below in section~\ref{sec:finite} (and in closed form for Gaussian switching in subsection~\ref{subsec:finite_gauss}).

Using \eqref{eq:trajectory_param}--\eqref{eq:dtau_lightcone}, substituting \eqref{eq:unruh_modes} into \eqref{eq:dyson}, and converting the $\tau$-integral into a Mellin-type integral over $u$ (RTW) or $v$ (LTW), one finds that the derivative coupling contributes the characteristic on-shell factor $\omega/a$. The remaining Mellin integrals are evaluated distributionally using
\begin{align}
\int_{-\infty}^{0}du\,(-u)^{\,i\Sigma-1}
=\int_{0}^{\infty}dv\,v^{\,i\Sigma-1}
=2\pi\,\delta(\Sigma),
\label{eq:mellin_delta}
\end{align}
which enforces the on-shell identification of the Unruh label with the detector ratio. It is convenient to package the resulting single-branch coefficient into a compact amplitude $I(\omega,a)$, defined via the RTW creation amplitude,
\begin{align}
I(\omega,a)
= 2\pi g\,\frac{\omega}a\;
   \frac{e^{+\frac{\pi\omega}{2a}}}
        {\sqrt{8\pi\,\frac{\omega}a\,\sinh\Big(\pi\frac{\omega}a\Big)}}\;
   a^{\,i\frac{\omega}a}.
\label{eq:I-spectral}
\end{align}
We then define, for each branch,
\begin{align}
\mathcal{I}_k \equiv I(-\omega_k,a_k),
\qquad
\Lambda_k\equiv\frac{\omega_k}{a_k}.
\label{eq:def_Ik_Lk}
\end{align}
In the adiabatic limit, the LTW contribution is the complex conjugate of the RTW one, so the first-order state can be written in the compact form
\begin{align}
\ket{\Psi_f}
=
\sum_{k=1}^{2}\alpha_k
\Big[
\mathcal{I}_k\,A^\dagger_{-\Lambda_k}
+\mathcal{I}_k^{*}\,B^\dagger_{+\Lambda_k}
\Big]\ket{0_M}\ket{e}\otimes\ket{a_k},
\label{eq:final_state}
\end{align}
where we have set $\hbar=1$.

\subsection{Conditional (path-erasing) readout}

Because the ancilla path states $\ket{a_1}$ and $\ket{a_2}$ are orthogonal, the ancilla can store which-branch information. If one measures only the detector (excited vs.\ ground) and does not interrogate the ancilla, then the path is unobserved and one must sum probabilities over it. In that unconditional setting the excitation probability is an incoherent sum of the two branch contributions: the interference term is proportional to $\braket{a_1}{a_2}=0$ and therefore vanishes. Hence no phase dependence can appear unless the ancilla is measured in a way that removes the which-path record.

To access interference, we measure the ancilla in a superposition basis, analogous to recombining the arms of an interferometer and selecting an output port. Concretely, we introduce
\begin{align}
\ket{\pm}
=&\frac{\ket{a_1}\pm\ket{a_2}}{\sqrt{2}},
\nn\\
\ket{a_1}=&\frac{\ket{+}+\ket{-}}{\sqrt{2}},
\qquad
\ket{a_2}=\frac{\ket{+}-\ket{-}}{\sqrt{2}},
\label{eq:pm_basis}
\end{align}
and the postselection projectors
$\Pi_{e,\pm}=\ket{e}\bra{e}\otimes\ket{\pm}\bra{\pm}$.
Only in such conditional channels do the two branch amplitudes add coherently.

It is useful to rewrite the first-order state \eqref{eq:final_state} directly in the $\ket{\pm}$ basis. Define the branch emission ket (field $\otimes$ detector, with the detector excited)
\begin{align}
\ket{\Phi_k}
\equiv
\Big[
\mathcal{I}_k\,A^\dagger_{-\Lambda_k}
+\mathcal{I}_k^{*}\,B^\dagger_{+\Lambda_k}
\Big]\ket{0_M}\ket{e},
\qquad (k=1,2),
\label{eq:Phi_k_def}
\end{align}
so that $\ket{\Psi_f}=\sum_k \alpha_k\,\ket{\Phi_k}\otimes\ket{a_k}$.
Using \eqref{eq:pm_basis} gives the port decomposition
\begin{align}
\ket{\Psi_f}
=&
\alpha_1\,\ket{\Phi_1}\otimes\ket{a_1}
+\alpha_2\,\ket{\Phi_2}\otimes\ket{a_2}\nn\\
=&
\frac{1}{\sqrt{2}}
\Big(\alpha_1\ket{\Phi_1}+\alpha_2\ket{\Phi_2}\Big)\otimes\ket{+}
\nn\\
&
+\frac{1}{\sqrt{2}}
\Big(\alpha_1\ket{\Phi_1}-\alpha_2\ket{\Phi_2}\Big)\otimes\ket{-}.
\label{eq:Psi_f_pm_decomp}
\end{align}
Thus the $\ket{+}$ output carries the \emph{coherent sum} of the two branch amplitudes, while the $\ket{-}$ output carries their \emph{coherent difference}. Projecting onto the path-erasing port $(e,+)$ yields
\begin{align}
\Pi_{e,+}\ket{\Psi_f}
=
\frac{1}{\sqrt{2}}
\Big(\alpha_1\ket{\Phi_1}+\alpha_2\ket{\Phi_2}\Big)\otimes\ket{+},
\label{eq:postselect_full_plus}
\end{align}
and similarly for the orthogonal port $(e,-)$,
\begin{align}
\Pi_{e,-}\ket{\Psi_f}
=
\frac{1}{\sqrt{2}}
\Big(\alpha_1\ket{\Phi_1}-\alpha_2\ket{\Phi_2}\Big)\otimes\ket{-}.
\label{eq:postselect_full_minus}
\end{align}
This is the operational content of WIT: suppression in the $(e,+)$ channel is a ``dark-port'' condition on the coherent sum.

Separating $\ket{\Phi_k}$ into its right- and left-moving pieces,
\begin{align}
\ket{\Phi_k}
=&
\ket{\Phi_k^{\mathrm{RTW}}}+\ket{\Phi_k^{\mathrm{LTW}}},
\nn\\
\ket{\Phi_k^{\mathrm{RTW}}}=&\mathcal{I}_k\,A^\dagger_{-\Lambda_k}\ket{0_M}\ket{e},
\nn\\
\ket{\Phi_k^{\mathrm{LTW}}}=&\mathcal{I}_k^{*}\,B^\dagger_{+\Lambda_k}\ket{0_M}\ket{e},
\label{eq:Phi_k_split}
\end{align}
one finds that projecting \eqref{eq:Psi_f_pm_decomp} onto $(e,+)$ gives the RTW and LTW contributions
\begin{align}
\ket{\Psi_{e,+}^{\mathrm{RTW}}}
&=
\frac{1}{\sqrt{2}}
\Big[
\alpha_1\,\mathcal{I}_1\,A^\dagger_{-\Lambda_1}
+\alpha_2\,\mathcal{I}_2\,A^\dagger_{-\Lambda_2}
\Big]\ket{0_M}\ket{e},
\label{eq:postselect_RTW_pre_match}
\\
\ket{\Psi_{e,+}^{\mathrm{LTW}}}
&=
\frac{1}{\sqrt{2}}
\Big[
\alpha_1\,\mathcal{I}_1^{*}\,B^\dagger_{+\Lambda_1}
+\alpha_2\,\mathcal{I}_2^{*}\,B^\dagger_{+\Lambda_2}
\Big]\ket{0_M}\ket{e}.
\label{eq:postselect_LTW_pre_match}
\end{align}

The one-particle states $A^\dagger_{-\Omega}\ket{0_M}$ at different Unruh labels are orthogonal, and likewise for $B^\dagger_{+\Omega}\ket{0_M}$. Therefore the two terms in \eqref{eq:postselect_RTW_pre_match} (and in \eqref{eq:postselect_LTW_pre_match}) lie in orthogonal subspaces unless they create the \emph{same} Unruh label. Exact cancellation of the postselected amplitude is therefore possible only if the two branches populate the same mode label, which requires
\begin{align}
\Lambda_1=\Lambda_2\equiv\Lambda,
\qquad
\Lambda_k:=\frac{\omega_k}{a_k}.
\label{match}
\end{align}
After imposing \eqref{match}, the conditional RTW and LTW components reduce to
\begin{align}
\ket{\Psi_{e,+}^{\mathrm{RTW}}}
&=
\frac{1}{\sqrt{2}}
\Big(\alpha_1\,\mathcal{I}_1+\alpha_2\,\mathcal{I}_2\Big)\,
A^\dagger_{-\Lambda}\ket{0_M}\ket{e},
\label{eq:postselect_RTW_matched}
\\
\ket{\Psi_{e,+}^{\mathrm{LTW}}}
&=
\frac{1}{\sqrt{2}}
\Big(\alpha_1\,\mathcal{I}_1^{*}+\alpha_2\,\mathcal{I}_2^{*}\Big)\,
B^\dagger_{+\Lambda}\ket{0_M}\ket{e}.
\label{eq:postselect_LTW_matched}
\end{align}
Thus, suppression of the conditional $(e,+)$ port requires the coherent sums to vanish,
\begin{align}
\alpha_1\,\mathcal{I}_1+\alpha_2\,\mathcal{I}_2=&0,
\label{eq:RTW_cancel_cond}
\\
\alpha_1\,\mathcal{I}_1^{*}+\alpha_2\,\mathcal{I}_2^{*}=&0.
\label{eq:LTW_cancel_cond}
\end{align}

To make the phase condition transparent, note that under \eqref{match} the magnitudes of the two on-shell coefficients coincide up to the explicit acceleration-dependent phase contained in $I(-\omega,a)$, namely the factor $a^{\,i\omega/a}$ in \eqref{eq:I-spectral}. In particular, with $\Lambda=\omega_1/a_1=\omega_2/a_2$ one has
\begin{align}
\frac{\mathcal{I}_2}{\mathcal{I}_1}
=\Big(\frac{a_2}{a_1}\Big)^{-i\Lambda},
\label{eq:I_ratio_phase}
\end{align}
while the remaining real prefactors agree branch-by-branch once evaluated on shell.
Consequently, \eqref{eq:RTW_cancel_cond} yields
\begin{align}
\frac{\alpha_2}{\alpha_1}
&=-\left(\frac{a_1}{a_2}\right)^{-i\Lambda},
\qquad
\mathrm{RTW},
\label{eq:RTW_phase}
\end{align}
and \eqref{eq:LTW_cancel_cond} gives the conjugate condition
\begin{align}
\frac{\alpha_2}{\alpha_1}
&=-\left(\frac{a_1}{a_2}\right)^{+i\Lambda},
\qquad
\mathrm{LTW}.
\label{eq:LTW_phase}
\end{align}
Equations~\eqref{match}, \eqref{eq:RTW_phase}, and \eqref{eq:LTW_phase} summarize the amplitude-level cancellation rules for the conditional (path-erasing) readout used below.

Finally, suppression in the $(e,+)$ channel does not imply that excitation is impossible. Rather, it means that the coherent sum into the $\ket{+}$ output cancels at first order. The corresponding excitation amplitude is then redirected into the orthogonal ancilla output $(e,-)$, as is evident from \eqref{eq:Psi_f_pm_decomp} and \eqref{eq:postselect_full_minus}.

\subsection{EIT mapping}

The conditional (path-erasing) readout isolates a two-path interference problem that is formally identical to the interference mechanism in a standard $\Lambda$-type EIT system. In EIT language, the detector has two long-lived ``ground'' configurations $\ket{g_1}$ and $\ket{g_2}$ (here distinguished by the branch ancilla) that are both coupled to a common excited state $\ket{e}$. The postselection onto $\ket{+}$ selects the coherent superposition of the two excitation pathways $\ket{g_k}\to\ket{e}$. In the present setting there is no external control field; rather, the effective ``couplings'' are the first-order branch excitation amplitudes. In the adiabatic limit these are proportional to $\mathcal{I}_k$ in the right-moving sector and to $\mathcal{I}_k^{*}$ in the left-moving sector.

It is useful to parametrize the branch weights by a single relative phase,
\begin{align}
\alpha_1
&=r\,e^{+i\theta/2},\qquad
\alpha_2=r\,e^{-i\theta/2},
\label{eq:alpha_param}
\end{align}
with $r$ fixed by normalization. Then
\begin{align}
\frac{\alpha_2}{\alpha_1}=e^{-i\theta}.
\label{eq:alpha_ratio_theta}
\end{align}
Using  \eqref{eq:RTW_phase} we have
\begin{align}
e^{-i\theta}
&=
-e^{-i\Lambda\ln(a_1/a_2)},
\end{align}
and hence the explicit RTW phase settings
\begin{align}
\theta
&=(2m+1)\pi+\Lambda\ln\!\left(\frac{a_1}{a_2}\right),
\qquad m\in\mathbb{Z}.
\label{eq:RTW_theta}
\end{align}
This derivation also makes transparent why only a \emph{single} tunable phase is needed: once the matching condition \eqref{match} holds, the on-shell magnitudes satisfy $|\mathcal{I}_1|=|\mathcal{I}_2|$, and the branch dependence of $\mathcal{I}_2/\mathcal{I}_1$ reduces entirely to the acceleration-dependent phase in \eqref{eq:I_ratio_phase}. The LTW sector is the complex conjugate channel, yielding the conjugate phase rule from \eqref{eq:LTW_phase},
\begin{align}
\theta
&=(2n+1)\pi-\Lambda\ln\!\left(\frac{a_1}{a_2}\right),
\qquad n\in\mathbb{Z}.
\label{eq:LTW_theta}
\end{align}
The terminology ``dark'' is used here in the standard interference sense: at the values of $\theta$ given above, the transition amplitude into the postselected $(e,+)$ output vanishes at first order in $g$, even though excitation still occurs in the orthogonal $(e,-)$ output.

Finally, if one wishes to suppress both RTW and LTW conditional channels simultaneously at the same setting of the relative phase, \eqref{eq:RTW_theta} and \eqref{eq:LTW_theta} must be compatible. This requires
\begin{align}
\Lambda\ln\!\left(\frac{a_1}{a_2}\right)
&=\pi k,
\qquad k\in\mathbb{Z}.
\label{eq:WIT_condition}
\end{align}
Equations~\eqref{match}–\eqref{eq:WIT_condition} provide the adiabatic cancellation rules used below.

\section{Plane-wave representation} \label{sec:plane_wave}

We adopt the same common--excited-state convention as above and re-express the first-order emission in the Minkowski plane-wave basis, which makes the interference structure explicit in terms of the inertial (Minkowski) frequency $\nu$.

We use light-cone coordinates $u=t-x$ and $v=t+x$. In the right Rindler wedge one has $u<0$ and $v>0$. The creation parts of the right- and left-moving field components admit the standard plane-wave expansions
\begin{align}
\Phi^-_{\mathrm{RTW}}(u)
=&\int_{0}^{\infty}\frac{d\nu}{\sqrt{4\pi\nu}}\;e^{i\nu u}\,a_\nu^\dagger,\nn\\
\Phi^-_{\mathrm{LTW}}(v)
=&\int_{0}^{\infty}\frac{d\nu}{\sqrt{4\pi\nu}}\;e^{i\nu v}\,b_\nu^\dagger,
\end{align}
where $\nu>0$ is the Minkowski frequency and the mode operators satisfy
$[a_\nu,a_{\nu'}^\dagger]=\delta(\nu-\nu')$ and $[b_\nu,b_{\nu'}^\dagger]=\delta(\nu-\nu')$, with $[a_\nu,b_{\nu'}]=[a_\nu,b_{\nu'}^\dagger]=0$.

The first-order state then reads
\begin{align}
\ket{\Psi_f}
=&-\frac{ig}{\hbar}\sum_{k=1}^{2}\alpha_k \nn\\
&\times
\int_{0}^{\infty}\frac{d\nu}{\sqrt{4\pi\nu}}\,
\Big[\mathcal{I}_u^{(k)}(\nu)\,a_\nu^\dagger+\mathcal{I}_v^{(k)}(\nu)\,b_\nu^\dagger\Big]\ket{0_M}\ket{e},
\label{eq:pw_final_state}
\end{align}
with branch-dependent light-cone integrals
\begin{align}
\mathcal{I}_u^{(k)}(\nu)
&=\int_{-\infty}^{0}du\,
\frac{\partial}{\partial u}\Big[e^{i\nu u}\,(-a_k u)^{-i\omega_k/a_k}\Big],\nn\\
\mathcal{I}_v^{(k)}(\nu)
&=\int_{0}^{\infty}dv\,
\frac{\partial}{\partial v}\Big[e^{i\nu v}\,(a_k v)^{+i\omega_k/a_k}\Big].\nn
\end{align}
Evaluating these oscillatory integrals with the standard $i\epsilon$ prescription (equivalently, by contour deformation and analytic continuation) gives
\begin{align}
\mathcal{I}_u^{(k)}(\nu)
&=\nu^{+i\omega_k/a_k}\,a_k^{-i\omega_k/a_k}\,
e^{-\frac{\pi\omega_k}{2a_k}}\,
\Gamma\Big(1-\frac{i\omega_k}{a_k}\Big),\\
\mathcal{I}_v^{(k)}(\nu)
&=-\,\nu^{-i\omega_k/a_k}\,a_k^{+i\omega_k/a_k}\,
e^{-\frac{\pi\omega_k}{2a_k}}\,
\Gamma\Big(1+\frac{i\omega_k}{a_k}\Big).
\end{align}

The mode operators admit frequency-dependent rephasings,
$a_\nu\to e^{i\phi(\nu)}a_\nu$ and $a_\nu^\dagger\to e^{-i\phi(\nu)}a_\nu^\dagger$ (similarly for $b_\nu$), without affecting commutators. However, this is a single phase convention choice for each $\nu$, independent of the detector branch. Therefore the branch factors $\nu^{\pm i\omega_k/a_k}$ can be absorbed into such a rephasing \emph{only when they coincide across branches}.

Equivalently, since the one-particle states $a_\nu^\dagger\ket{0_M}$ at different $\nu$ are orthogonal (and hence linearly independent), cancellation of the RTW (or LTW) contribution at the operator level requires the coefficient multiplying $a_\nu^\dagger$ (or $b_\nu^\dagger$) to vanish \emph{for each} $\nu$. This enforces the matching condition
\[
\frac{\omega_1}{a_1}=\frac{\omega_2}{a_2}\equiv\Lambda.
\]

Under this matching, the $\nu$-dependence becomes common and can be removed by a single rephasing, reducing the RTW coefficient to
\begin{align}
\alpha_1\,a_1^{-i\Lambda}\,e^{-\frac{\pi\Lambda}{2}}\Gamma(1-i\Lambda)
+\alpha_2\,a_2^{-i\Lambda}\,e^{-\frac{\pi\Lambda}{2}}\Gamma(1-i\Lambda)=0,
\end{align}
which yields exactly the phase condition found in the Unruh-mode analysis,
\begin{align}
\frac{\alpha_2}{\alpha_1}=-\Big(\frac{a_1}{a_2}\Big)^{-i\Lambda}.
\end{align}
The LTW sector follows identically, with the conjugate condition
$\alpha_2/\alpha_1=-(a_1/a_2)^{+i\Lambda}$, using
$\big[\Gamma(1+ix)\big]^*=\Gamma(1-ix)$ for real $x$.

\begin{figure*}[t!]
\centering
\includegraphics[width=\columnwidth]{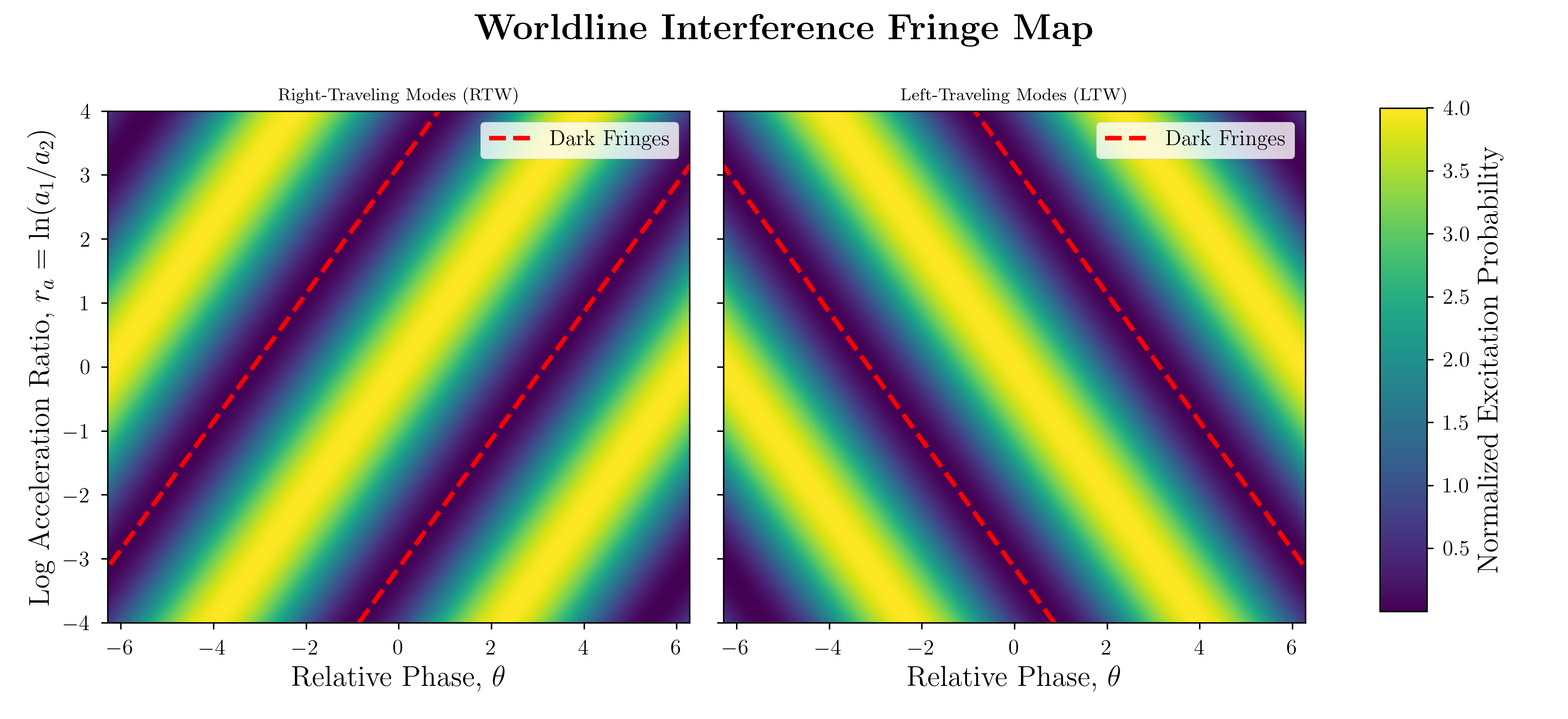}
\vspace{0.5cm}
\includegraphics[width=\columnwidth]{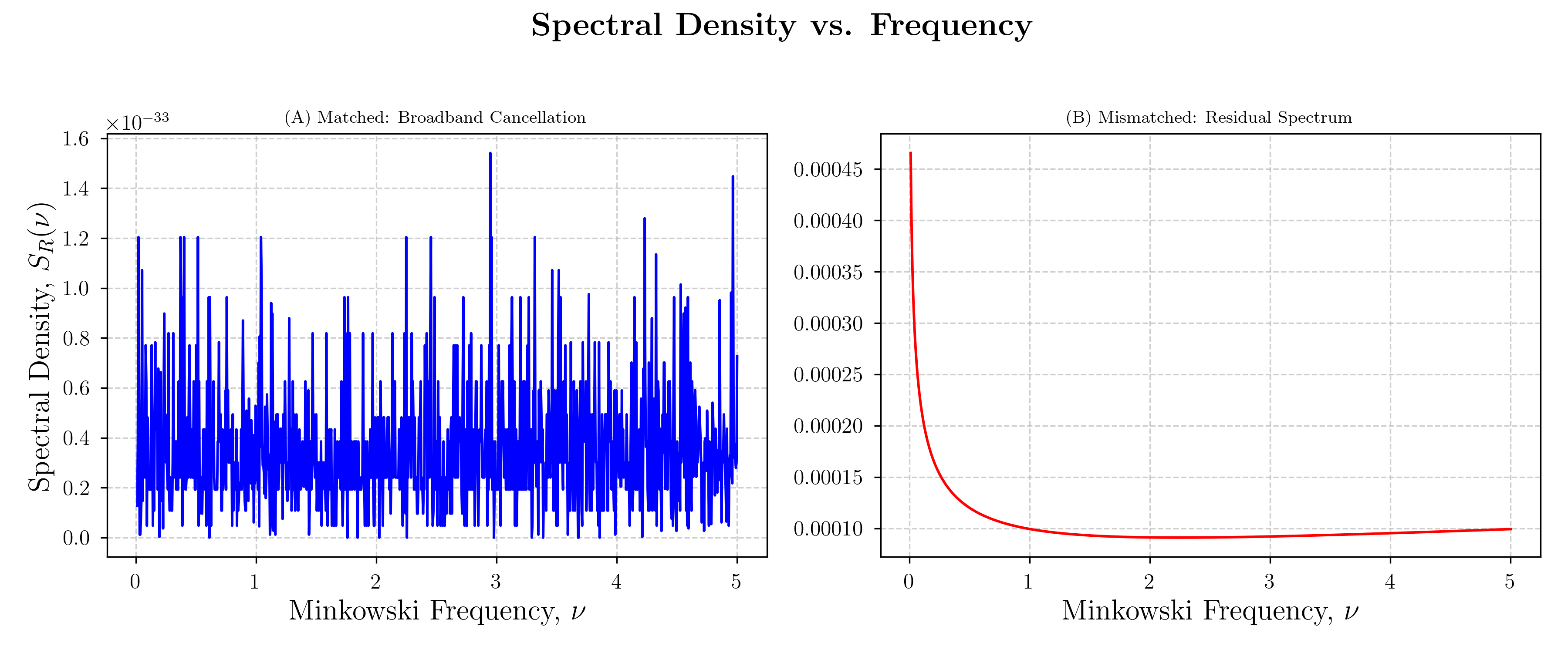}
\caption{
Conditional interference and spectral cross-check.
\textbf{(Left)} Color map (``heatmap'') of the normalized conditional excitation probability in the postselected output $(e,+)$ as a function of the relative branch phase $\theta$ and the log-acceleration ratio $r_a\equiv\ln(a_1/a_2)$.
Dark regions indicate suppression (destructive interference), while bright regions indicate enhancement (constructive interference).
The dashed curves mark the analytic cancellation loci predicted by Eqs.~\eqref{eq:RTW_theta}--\eqref{eq:WIT_condition}.
\textbf{(Right)} Corresponding emitted-radiation spectrum (plotted versus Minkowski frequency $\nu$) for two representative parameter choices.
The ``on-fringe'' choice is taken on a suppression curve in the left panel and yields strong reduction of the spectrum over the plotted range.
The ``off-fringe'' choice includes a small ratio mismatch $\Delta\Lambda\neq 0$ (equivalently $\omega_1/a_1\neq \omega_2/a_2$), so the two branch contributions populate different frequency/Unruh-label bands and do not cancel perfectly, leaving a residual structured spectrum.
}
\label{fig:combined_verification}
\end{figure*}

This establishes consistency between the Unruh-mode and plane-wave representations and clarifies that the $\nu$-phase factors can be dropped only \emph{after} the matching is imposed. This operator-level analysis matches the spectral verification in the right panel of Fig.~\ref{fig:combined_verification}, where exact matching collapses the spectrum and small detunings restore structured emission.

\begin{figure}[t!]
\centering
\includegraphics[width=\columnwidth]{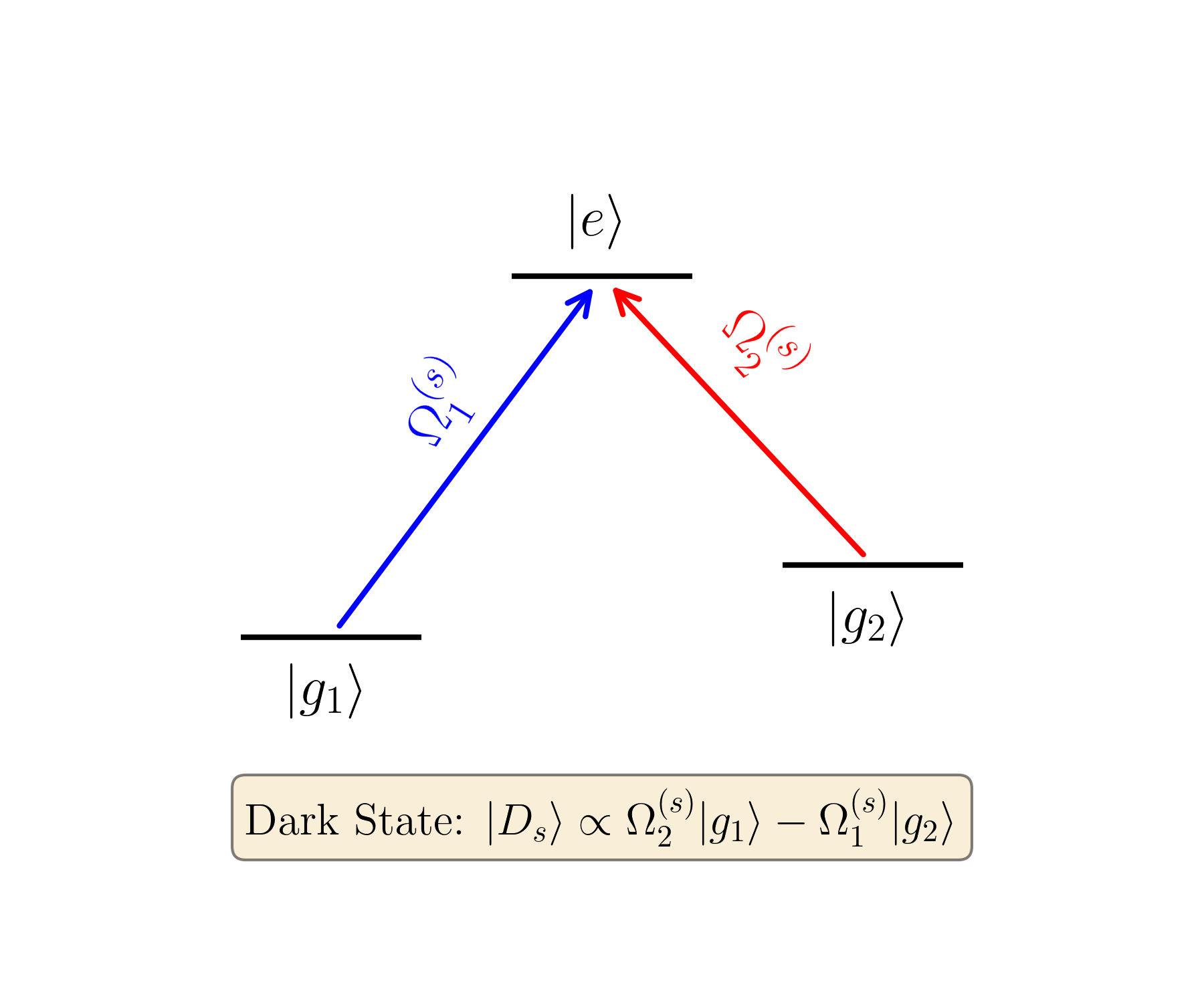}
\caption{
Analogy to a three-level $\Lambda$ system.
The branch-conditioned detector configurations $\ket{g_1}$ and $\ket{g_2}$ play the role of two long-lived ``ground'' states coupled to a common excited state $\ket{e}$.
The two excitation pathways (one from each branch) interfere in the \emph{conditional} (path-erasing) readout, i.e., after projecting the branch ancilla onto $\ket{+}=(\ket{a_1}+\ket{a_2})/\sqrt{2}$.
In this language the effective coupling strengths are proportional to the first-order branch excitation amplitudes (RTW: $\mathcal{I}_k$, LTW: $\mathcal{I}_k^{*}$ in the adiabatic limit).
For the relative phase settings that satisfy the cancellation rules in Eqs.~\eqref{match}--\eqref{eq:WIT_condition}, the transition amplitude into the postselected $(e,+)$ channel vanishes at first order (``dark'' output), while excitation can still appear in the orthogonal $(e,-)$ outcome.
}
\label{fig:eit_schematic}
\end{figure}

\section{Finite interaction time and natural bandwidth}
\label{sec:finite}

The adiabatic formulas in Eqs.~\eqref{eq:final_state} and \eqref{match} are obtained when the proper-time integral acts as a sharp selector of the Unruh label. For a finite interaction duration this selector broadens into a spectral envelope determined by the switching profile. In this section we present, first, an exact finite-time result for Gaussian switching (kept fully explicit at first order), and then we extract the corresponding natural bandwidth interpretation and tolerance window in the dimensionless ratio $\omega/a$.

\subsection{Gaussian switching: exact first-order amplitude and tolerance window}
\label{subsec:finite_gauss}

For a Gaussian switch \cite{Azizi2025TPE_Switch} one can keep the first Dyson term exactly and obtain a closed-form, frequency-resolved first-order excitation amplitude. The resulting ``tolerance window'' quantifies when the two branch wavepackets overlap in Unruh-label space and can therefore interfere in the conditional (path-erasing) readout.

\subsubsection*{Gaussian gate and its Fourier profile}

We take a smooth Gaussian window of characteristic duration $T$,
\begin{align}
s_T(\tau)
=
\frac{1}{T\sqrt{2\pi}}\exp\!\Big(-\frac{\tau^2}{2T^2}\Big),
\label{eq:gauss_switch_def}
\end{align}
so that $\int d\tau\, s_T(\tau)=1$. This choice corresponds to a \emph{fixed pulse area} protocol: increasing $T$ spreads the interaction in time while keeping $\int s_T$ fixed. Its proper-time Fourier transform is exact:
\begin{align}
\tilde{s}_T(\Sigma)
\equiv
\int_{-\infty}^{+\infty}d\tau\,s_T(\tau)\,e^{i\Sigma\tau}
=
\exp\!\Big(-\frac{T^2\Sigma^2}{2}\Big),
\label{eq:gauss_switch_FT}
\end{align}
where $\Sigma$ has dimensions of frequency.

For comparison with the standard ``long-time $\to$ delta selector'' intuition (appropriate when one keeps the coupling strength fixed while extending the interaction time), it is convenient to introduce the unnormalized gate
$\chi_T(\tau)\equiv T\sqrt{2\pi}\,s_T(\tau)=\exp(-\tau^2/2T^2)$, for which
$\tilde{\chi}_T(\Sigma)=T\sqrt{2\pi}\,e^{-T^2\Sigma^2/2}$ and hence
\begin{align}
\lim_{T\to\infty}\frac{1}{2\pi}\tilde{\chi}_T(\Sigma)
=
\delta(\Sigma)
\qquad \text{(distributionally)}.
\label{eq:gauss_delta_sequence}
\end{align}
Thus, the ``delta selector'' limit corresponds either to working with $\chi_T$ or, equivalently, to rescaling the effective coupling by the factor $T\sqrt{2\pi}$ when using the unit-area gate $s_T$. In what follows we keep $s_T$ for definiteness; the spectral widths and overlap criteria are unaffected by this overall normalization choice.

\subsubsection*{Exact first-order Unruh-mode amplitude at finite time}

Starting from the first Dyson term and retaining the finite $\tau$-integral, one obtains the exact frequency-resolved excited components in the Unruh basis (see Appendix~\ref{app:gauss_derivation}):
\begin{align}
\ket{\Psi_f^{\mathrm{RTW}}}
=&\ g \sum_{k=1}^{2}\alpha_k
\int_{-\infty}^{+\infty} d\Omega\;
(a_k\Omega)\,f(-\Omega)\,a_k^{\,i\Omega} \label{eq:gauss_state_RTW}\\
&\times
\exp\!\Big[-\frac{T^2}{2}(\omega_k+a_k\Omega)^2\Big]
A^\dagger_{\Omega}\ket{0_M}\ket{e}\otimes\ket{a_k},
\nn
\end{align}
and
\begin{align}
\ket{\Psi_f^{\mathrm{LTW}}}
=&\ g \sum_{k=1}^{2}\alpha_k
\int_{-\infty}^{+\infty} d\Omega\;
(a_k\Omega)\,f(\Omega)\,a_k^{\,i\Omega} \label{eq:gauss_state_LTW}\\
&\times
\exp\Big[-\frac{T^2}{2}(\omega_k-a_k\Omega)^2\Big]
B^\dagger_{\Omega}\ket{0_M}\ket{e}\otimes\ket{a_k}.
\nn
\end{align}

Here we adopt the same Unruh-mode convention as above: we allow the Unruh label $\Omega\in\mathbb R$ in the creation-operator sector, so that the RTW excitation produced by the derivative coupling is supported near negative $\Omega$. Defining $\Lambda_k:=\omega_k/a_k>0$, Eqs.~\eqref{eq:gauss_state_RTW}--\eqref{eq:gauss_state_LTW} show that each branch populates a narrow band centered at
$\Omega\simeq-\Lambda_k$ (RTW) and $\Omega\simeq+\Lambda_k$ (LTW), with characteristic width
\begin{align}
\Delta\Omega_k
\sim \frac{1}{a_k T},
\label{eq:gauss_bandwidth}
\end{align}
up to an $\mathcal{O}(1)$ definition-dependent factor. We assume throughout that $\Lambda_k$ is not parametrically small, so that the mode prefactors $(a_k\Omega)f(\pm\Omega)$ vary slowly across each packet.

\subsubsection*{Path-erasing readout and overlap-controlled interference}

In the conditional (path-erasing) port $\ket{+}=(\ket{a_1}+\ket{a_2})/\sqrt{2}$, the RTW excited component takes the form
\begin{align}
\ket{\Psi_{e,+}^{\mathrm{RTW}}}
\propto\ &
\int d\Omega\;\Big(\alpha_1\,\mathcal{F}_1^{\mathrm{R}}(\Omega)+\alpha_2\,\mathcal{F}_2^{\mathrm{R}}(\Omega)\Big)\nn\\
&\times A^\dagger_{\Omega}\ket{0_M}\ket{e},
\label{eq:gauss_postselect_RTW_general}
\end{align}
with
$\mathcal{F}_k^{\mathrm{R}}(\Omega)=(a_k\Omega)\,f(-\Omega)\,a_k^{i\Omega}\,
\exp\!\big[-\frac{T^2}{2}(\omega_k+a_k\Omega)^2\big]$.
Because $A^\dagger_{\Omega}\ket{0_M}$ at different $\Omega$ are orthogonal, interference in the conditional probability is controlled by the spectral overlap of the two branch wavepackets in $\Omega$-space.

In the regime $a_kT\gg 1$, the Gaussians are narrow in $\Omega$ and one may evaluate the slowly varying prefactors at the packet centers. Defining the RTW Gaussian envelopes
\begin{align}
G_k^{\mathrm{R}}(\Omega)
\equiv
\exp\!\Big[-\frac{(a_kT)^2}{2}\,(\Omega+\Lambda_k)^2\Big],
\label{eq:GaussEnvelopeDef}
\end{align}
the spectral amplitude is well-approximated by a sum of two narrow packets,
\begin{align}
\alpha_1\,\mathcal{F}_1^{\mathrm{R}}(\Omega)+\alpha_2\,\mathcal{F}_2^{\mathrm{R}}(\Omega)
\ \simeq\ 
\alpha_1\,\mathcal{I}_1\,G_1^{\mathrm{R}}(\Omega)
+\alpha_2\,\mathcal{I}_2\,G_2^{\mathrm{R}}(\Omega),
\label{eq:gauss_packet_sum}
\end{align}
where $\mathcal{I}_k$ denotes the corresponding on-shell coefficient obtained by evaluating the prefactors at $\Omega=-\Lambda_k$. This identification matches the adiabatic coefficient used in \eqref{eq:final_state} once the same switching normalization is adopted (equivalently, using $\chi_T$ so that $\tilde\chi_T\to 2\pi\delta$), or upon the corresponding rescaling of the effective coupling.

A further condition is required for the interference cross-term to be governed primarily by Gaussian overlap: the relative phase factor
$a_1^{i\Omega}a_2^{-i\Omega}=e^{i\Omega\ln(a_1/a_2)}$ must not oscillate appreciably across the overlap width. A sufficient criterion is
\begin{align}
\Delta\Omega_{\mathrm{eff}}\;|\ln(a_1/a_2)|\ll 1,
\qquad
\Delta\Omega_{\mathrm{eff}}\sim \frac{1}{a_{\mathrm{eff}}T},
\label{eq:phase_overlap_condition}
\end{align}
with $a_{\mathrm{eff}}$ any acceleration scale comparable to $a_1,a_2$ (e.g.\ $a_{\mathrm{eff}}=\max\{a_1,a_2\}$).
When \eqref{eq:phase_overlap_condition} holds, the envelope overlap controls the interference visibility.

\subsubsection*{\texorpdfstring{Finite-time tolerance window in $\omega/a$}{}}

A necessary condition for appreciable interference is that the two Gaussian bands overlap in $\Omega$. The overlap integral
$\int d\Omega\,G_1^{\mathrm{R}}(\Omega)\,G_2^{\mathrm{R}}(\Omega)$ evaluates exactly to
\begin{align}
\int_{-\infty}^{+\infty} d\Omega\;G_1^{\mathrm{R}}(\Omega)\,G_2^{\mathrm{R}}(\Omega)
=&
\sqrt{\frac{2\pi}{T^2(a_1^2+a_2^2)}} \label{eq:GaussOverlapIntegral}\\
&\times
\exp\!\Bigg[-\frac{a_1^2a_2^2T^2}{2(a_1^2+a_2^2)}\,(\Delta\Lambda)^2\Bigg],
\nn
\end{align}
where
$\Delta\Lambda\equiv\Lambda_1-\Lambda_2=\omega_1/a_1-\omega_2/a_2$.
Thus, overlap (and hence interference) is exponentially suppressed once
$\frac{a_1^2a_2^2T^2}{2(a_1^2+a_2^2)}(\Delta\Lambda)^2\gg 1$. Equivalently, a conservative Gaussian tolerance criterion is
\begin{align}
\bigg|\frac{\omega_1}{a_1}-\frac{\omega_2}{a_2}\bigg|
\ \lesssim\
\frac{\sqrt{a_1^2+a_2^2}}{a_1a_2}\,\frac{1}{T},
\label{eq:finiteTband_gauss}
\end{align}
which scales as $|\Delta\Lambda|\sim 1/(aT)$ with $a$ any acceleration scale comparable to $a_1,a_2$.

\section{Conclusion} \label{sec:conclusion}

We have analyzed the leading-order response of a single Unruh--DeWitt detector whose trajectory is coherently correlated with a two-state path ancilla, so that the interaction proceeds along one of two uniformly accelerated branches while the final ancilla measurement can either retain or erase which-branch information. If the ancilla is not read out, the orthogonality $\braket{a_1}{a_2}=0$ enforces an incoherent sum over branches and removes any dependence on the relative phase of the branch weights. By contrast, conditioning on an ancilla superposition output (the path-erasing port) makes the excitation amplitude a coherent sum of the two branch amplitudes.

In that conditional channel, amplitude-level cancellation requires the two branches to populate the same one-particle subspace. In the Unruh-mode formulation this appears as the matching condition
\begin{align}
\frac{\omega_1}{a_1}
=&\ \frac{\omega_2}{a_2}
\equiv \Lambda,
\end{align}
after which the remaining branch dependence at first order is a relative phase carried by the acceleration-dependent factors in the single-branch coefficients. The cancellation phase rules in the postselected $(e,+)$ channel are
\begin{align}
\frac{\alpha_2}{\alpha_1}
=&\ -\left(\frac{a_1}{a_2}\right)^{-i\Lambda},
\qquad \mathrm{RTW},
\\
\frac{\alpha_2}{\alpha_1}
=&\ -\left(\frac{a_1}{a_2}\right)^{+i\Lambda},
\qquad \mathrm{LTW},
\end{align}
which are complex conjugates. Hence simultaneous suppression of both chiral contributions in the same conditional port is possible only at the discrete points where
$\Lambda\ln(a_1/a_2)=n\pi$ $(n\in\mathbb{Z})$; otherwise one may tune the phase to suppress RTW or LTW in a chosen port, but not both at once at leading order.

Finite interaction time replaces the sharp Unruh-label selection of the adiabatic limit by the Fourier profile of the switching window, so exact matching broadens into an overlap requirement in Unruh-label space. For Gaussian switching the first Dyson term can be evaluated in closed form, yielding narrow wavepackets centered near $\Omega\simeq\mp\,\omega_k/a_k$ with width $\Delta\Omega\sim 1/(aT)$, and therefore an explicit tolerance window $|\Delta(\omega/a)|\sim 1/(aT)$ controlling interference visibility (Sec.~\ref{subsec:finite_gauss}). A related interference-based suppression mechanism in a different architecture is presented in Ref.~\cite{Azizi2026SavingUnruh}.

\section*{Acknowledgments}

I am grateful to Marlan Scully, Philip Stamp, Bill Unruh, and Charles Wallace for stimulating and insightful discussions. This work was supported by the Robert A. Welch Foundation (Grant No. A-1261).

\appendix
\section{\texorpdfstring{Gaussian switching: derivation of Eqs.~\eqref{eq:gauss_state_RTW} and \eqref{eq:gauss_state_LTW}}{}}
\label{app:gauss_derivation}

We derive the exact Unruh-basis excited components for a Gaussian switching gate by evaluating the first Dyson term at finite interaction time. We set $\hbar=1$ throughout this appendix.

\begin{align}
\ket{\Psi_f}
=&-i\int_{-\infty}^{+\infty} d\tau\,H_I(\tau)\ket{\Psi_i}.
\end{align}
Acting on $\ket{\Psi_i}$, only the raising part of the monopole contributes to excitation,
$m(\tau)\ket{g}_{a_k}=e^{+i\omega_k\tau}\ket{e}\otimes\ket{a_k}$, so the excited component reads
\begin{align}
\ket{\Psi_f^{(e)}}
=&-ig\sum_{k=1}^{2}\alpha_k
\int_{-\infty}^{+\infty} d\tau\, s_T(\tau)\,e^{+i\omega_k\tau}\nn\\
&\times
\frac{d}{d\tau}\Big[\Phi_{\mathrm{RTW}}\!\big(u_k(\tau)\big)
+\Phi_{\mathrm{LTW}}\!\big(v_k(\tau)\big)\Big]\nn\\
&
\ket{0_M}\ket{e}\otimes\ket{a_k},
\label{eq:app_dyson_excited}
\end{align}
where we take a normalized Gaussian window
\begin{align}
s_T(\tau)
=&\ \frac{1}{\sqrt{2\pi}\,T}\,
\exp\!\Big[-\frac{\tau^2}{2T^2}\Big].
\label{eq:app_gauss_switch}
\end{align}

We begin with the RTW contribution. Using the Unruh-mode expansion (creation part)
\begin{align}
\Phi_{\mathrm{RTW}}^{-}(u)
=&\ \int_{-\infty}^{+\infty} d\Omega\;
f(-\Omega)\,(-u)^{-i\Omega}\,A^\dagger_{\Omega},
\end{align}
and the uniformly accelerated right-wedge trajectory
$-a_k u_k(\tau)=e^{-a_k\tau}$, one has
\begin{align}
(-u_k(\tau))^{-i\Omega}
=&\ a_k^{\,i\Omega}\,e^{+i\Omega a_k\tau}.
\label{eq:app_u_mellin_factor}
\end{align}
Moreover, along branch $k$,
$\frac{d}{d\tau}=(-a_k u_k)\partial_{u_k}$, so
\begin{align}
\frac{d}{d\tau}\,(-u_k)^{-i\Omega}
=&\ i(a_k\Omega)\,(-u_k)^{-i\Omega}
\nn\\
=&\ i(a_k\Omega)\,a_k^{\,i\Omega}\,e^{+i\Omega a_k\tau}.
\label{eq:app_d_tau_on_u_mode}
\end{align}
Substituting \eqref{eq:app_u_mellin_factor}--\eqref{eq:app_d_tau_on_u_mode} into \eqref{eq:app_dyson_excited} and collecting the RTW creation operator piece gives
\begin{align}
\ket{\Psi_f^{\mathrm{RTW}}}
=&-ig\sum_{k=1}^{2}\alpha_k
\int_{-\infty}^{+\infty} d\Omega\;
f(-\Omega)\,i(a_k\Omega)\,a_k^{\,i\Omega}\nn\\
&\times
\int_{-\infty}^{+\infty} d\tau\,
s_T(\tau)\,e^{i(\omega_k+a_k\Omega)\tau} \nn\\
&\times
A^\dagger_{\Omega}\ket{0_M}\ket{e}\otimes\ket{a_k}.
\label{eq:app_RTW_before_gauss_int}
\end{align}
The remaining $\tau$-integral is the Fourier transform of a Gaussian:
\begin{align}
\int_{-\infty}^{+\infty} d\tau\,
s_T(\tau)\,e^{i\Sigma\tau}
=&\ \exp\!\Big[-\frac{T^2}{2}\,\Sigma^2\Big].
\label{eq:app_gauss_fourier}
\end{align}
Applying \eqref{eq:app_gauss_fourier} with $\Sigma=\omega_k+a_k\Omega$ yields Eq.~\eqref{eq:gauss_state_RTW}.

For the LTW contribution, we use
\begin{align}
\Phi_{\mathrm{LTW}}^{-}(v)
=&\ \int_{-\infty}^{+\infty} d\Omega\;
f(\Omega)\,v^{-i\Omega}\,B^\dagger_{\Omega},
\end{align}
together with $a_k v_k(\tau)=e^{+a_k\tau}$, which implies
\begin{align}
v_k(\tau)^{-i\Omega}
=&\ a_k^{\,i\Omega}\,e^{-i\Omega a_k\tau}.
\label{eq:app_v_mellin_factor}
\end{align}
Since $\frac{d}{d\tau}=(a_k v_k)\partial_{v_k}$ on branch $k$, we obtain
\begin{align}
\frac{d}{d\tau}\,v_k^{-i\Omega}
=&\ -i(a_k\Omega)\,v_k^{-i\Omega}
\nn\\
=&\ -i(a_k\Omega)\,a_k^{\,i\Omega}\,e^{-i\Omega a_k\tau}.
\label{eq:app_d_tau_on_v_mode}
\end{align}
Substituting \eqref{eq:app_v_mellin_factor}--\eqref{eq:app_d_tau_on_v_mode} into \eqref{eq:app_dyson_excited} gives
\begin{align}
\ket{\Psi_f^{\mathrm{LTW}}}
=&-ig\sum_{k=1}^{2}\alpha_k
\int_{-\infty}^{+\infty} d\Omega\;
f(\Omega)\,\big[-i(a_k\Omega)\big]\,a_k^{\,i\Omega}\nn\\
&\times
\int_{-\infty}^{+\infty} d\tau\,
s_T(\tau)\,e^{i(\omega_k-a_k\Omega)\tau} \nn\\
&\times
B^\dagger_{\Omega}\ket{0_M}\ket{e}\otimes\ket{a_k}.
\label{eq:app_LTW_before_gauss_int}
\end{align}
Using \eqref{eq:app_gauss_fourier} with $\Sigma=\omega_k-a_k\Omega$ yields the Gaussian envelope
$\exp\!\big[-\frac{T^2}{2}(\omega_k-a_k\Omega)^2\big]$. The overall phase of the LTW creation sector may be fixed (by a constant rephasing of $B^\dagger_{\Omega}$) so that the adiabatic LTW amplitude coincides with the complex conjugate of the RTW one; with this convention the result is precisely Eq.~\eqref{eq:gauss_state_LTW}.

\bibliographystyle{apsrev4-2} 
\bibliography{UnruhRef}
\end{document}